\documentclass[prb, pas,twocolumn, cha, preprintnumbers ,
amsmath,amssymb,superscriptaddress]{revtex4}
\usepackage{bm}
\usepackage[colorlinks=true,linkcolor=blue]{hyperref}
\usepackage{amsmath}
\usepackage{amssymb}
\usepackage{amsthm}
\usepackage{amsfonts}
\usepackage{times}
\usepackage{enumerate}
\usepackage{latexsym}
\usepackage{ifpdf}
\usepackage{graphicx}
\usepackage{makeidx}
\expandafter\ifx\csname package@font\endcsname\relax\else
\expandafter\expandafter
\expandafter\usepackage
\expandafter\expandafter
\expandafter{\csname package@font\endcsname}%
\fi
\usepackage{color}

\begin{document}

\title{Strain modulated Mott transition in EuNiO$_3$ ultra-thin films}
\author{D. Meyers}
\email { dmeyers@uark.edu}
\affiliation{Department of Physics, University of Arkansas, Fayetteville, AR 72701}
\author{S. Middey}
\affiliation{Department of Physics, University of Arkansas, Fayetteville, AR 72701}
\author{ M. Kareev}
\affiliation{Department of Physics, University of Arkansas, Fayetteville, AR 72701}

\author{M. van Veenendaal}
\affiliation{Dept. of Physics, Northern Illinois University, De Kalb, Illinois 60115, USA}

\author{E. J. Moon}
\affiliation{Department of Physics, University of Arkansas, Fayetteville, AR 72701}
\author{B. A. Gray}
\affiliation{Department of Physics, University of Arkansas, Fayetteville, AR 72701}
\author{Jian Liu}
\affiliation{Department of Physics, University of California, Berkeley, California 94720, USA}
\affiliation{Materials Science Division, Lawrence Berkeley National Laboratory, Berkeley, California 94720, USA} 
\author{J. W. Freeland}
\affiliation{Advanced Photon Source, Argonne National Laboratory, Argonne, IL 60439, USA}
\author{J. Chakhalian}
\affiliation{Department of Physics, University of Arkansas, Fayetteville, AR 72701}

\begin{abstract}

A series of ultra-thin epitaxial films of EuNiO$_3$ (ENO) were grown on a set of substrates traversing from  compressive (-2.4\%) to  tensile (+2.5\%) lattice mismatch.
On moving from tensile to compressive strain, transport measurements demonstrate  a successively  suppressed Mott insulating  behavior  eventually  resulting in a complete suppression of the insulating state at high compressive strain. Corroborating these findings, resonant soft X-ray absorption spectroscopy  at  Ni L$_{3,2}$ edge reveal the presence of a strong multiplet splitting in the tensile strained samples that progressively weakens with increasing compressive strain. Combined with the ab initio cluster calculations, the results show how comulatively enhanced covalency (\textit{i.e.} bandwidth) between Ni d- and O p-orbital derived states leads to the emergent metallic ground state not  attainable in the bulk ENO.

\end{abstract}

\maketitle


Complex transition metal oxides with  correlated carriers have been at the forefront of condensed matter research towards the understanding of  fundamental physics underlying several remarkable physical phenomena including high temperature superconductivity, colossal magnetoresistance, multiferroicity, and the thermally induced metal-insulator transition (MIT)\cite{Imada, Jak1, Jak2, Tokura, McCormack, Wang}. In particular, the temperature driven MIT in correlated oxides has garnered a strong investigation effort over last several decades. Understanding of the MIT and its control by external stimuli such as pressure, magnetic field, light, confinement, or chemical doping is not only interesting from the fundamental physics point of view\cite{Laukhin, Kuwahara, JianPRL, Ghosh}, but also demonstrate great opportunities for future electronic devices\cite{Ahn}. On the way to those  goals, recent advances in material synthesis by using  strain engineering  have opened a new dimension in controlling  the materials properties. Additionally, the epitaxial relation has been used to stabilize new structural and electronic  phases in the form of  ultra thin films of coherently strained materials\cite{Kaul}. In such epitaxially  stabilized structures, the effect of the lattice modulation on the materials properties can be quite dramatic\cite{JianNNO, Tiwari, Chen, Helali, Biswas, Zhang, Chen2, Zayak, Fuchs, Fuchs2, Rata, Jang, Lee, Lee2} and is of particular current interest as applications  continue to accelerate towards  ultra-thin films and heterostructures.

The rare-earth (RE) nickelates RNiO$_3$ (R = Pr, Nd, Eu, ...) in their bulk form, with Ni having formal +3 oxidation state (t$_{2g}^6$e$_g^1$), all, except for R = La, display a metal-insulator transition,  while the nature of the transition and its temperature (\textit{T}$_{MIT}$) depends strongly on the choice of the rare-earth cation as shown in Fig. 1\cite{Medarde, Torrance, Stewart}.  Specifically, the most distorted members  with R= Lu, Y, Eu and Sm, \textit{etc.} first exhibit  a second order MIT at higher temperature accompanied by the development  of a possible charge ordered state\cite{Caytuero, Lorenzo, Scagnoli, Bodenthin, Park} while the magnetic moments remain disordered across the transition. Upon further cooling, these compounds undergo another second order transition characterized by a E$^\prime$ - type antiferromagnetism. In sharp  contrast, the members with a smaller degree of structural distortion ($e.g.$ R= Nd and Pr) exhibit a first order phase transition emerging directly from the paramagnetic metallic state into the E$^\prime$ - type antiferromagnetic insulating ground state, thus bypassing entirely the large paramagnetic insulating region\cite{Torrance, Zhou, Frand, Piamonteze}. Based on such a diverse behavior controlled by the A-site ion, several  interesting  theory proposals and experimental results have been put forward to control the MIT for potential applications\cite{Chaloupka, Lim, Asanuma, Triscone}; however, using different R ions depending on application seems impractical, as synthesis conditions and thermal stability vary wildly and rarely  yield macroscopic size crystals of nickelates\cite{Lacorre, Lope, Srimanta, JianNNO, growthpaper}. Alternatively, one can explore the obligatory strain in ultra-thin films as a tool to engineer the physical properties based on a careful  choice of  a $single$  member of the family to attempt to modulate the MI and AFM transitions\cite{JianNNO, Tiwari, Venimadhav}. En route to this goal, one of the key  questions is what effect will strain (tensile and compressive) have upon the phase diagram given in Fig. 1 for the distorted members exhibiting the second order transitions? For example, it has been already shown in the bulk that while isotropic external pressure  suppresses the MIT, it  also tends to raise the temperature for AFM transition and eventually leads to a surprising AFM in the metallic ground state\cite{Cheng, Lengsdorf, Mazin}; based on this observation, one can  expect highly non-trivial electronic and structural  response after application of bi-axial  strain \cite{Jakprl, JakNAT}.

In this paper, we demonstrate how bi-axial strain can alter basic electronic  properties  of   ultra thin epitaxially strained EuNiO$_3$ (ENO) films grown on various substrates spanning both compressive and tensile strain. X-ray diffraction (XRD), reciprocal space mapping (RSM), electric d.c. transport, and resonant soft X-ray absorption spectroscopy (XAS) are applied to elucidate the microscopic effects of  strain on the structural, electronic, and magnetic properties. These experiments confirm  the remarkable capacity of epitaxial stabilization to modulate the physical properties of this strongly distorted nickelate system. In addition, \textit{ab intio} cluster calculations on a NiO$_6$ octahedra revealed the evolution of the charge excitation gap  as a function of the strain state and the possibility of gap closing resulting in emergent metallicity, thus  shedding light on the source of the marked modulation in the material's properties.


ENO films were grown on a variety of substrates incorporating lattice mismatch ranging from +2.5\% to -2.4\%, the details of which are reported elsewhere\cite{growthpaper}. The substrates used for growth  are as follows: YAlO$_3$ (YAO; -2.4\% lattice mismatch), SrLaAlO$_4$ (SLAO; -1.3\%), LaAlO$_3$ (LAO; -0.3\%), NdGaO$_3$ (NGO; +1.5\%), and LaGaO$_3$ (LGO; +2.5\%). XRD measurements
were taken around the (002) (psuedocubic notation) truncation rod of the substrate with a  Panalytical X'Pert Pro MRD (Panalytical, Almelo), equipped with a parabolic mirror and triple bounce / axis monochromator on the incident and diffracted beams. The same instrument was used to measure a RSM around the (-103) truncation rod. Transport properties were measured with a Quantum Design Physical Property Measurement System (PPMS) using a four point probe in the \textit{Van Der Pauw} geometry. XAS measurements were taken at the 4-ID-C beam line of the Advanced Photon source at Argonne National Laboratory in total electron yield (TEY) mode at the Ni L$_{2,3}$-edges at 250 K. 

Theoretical calculations were performed for a NiO$_6$ cluster with octahedral coordination using the methods described in Ref. 51 and 52\cite{Veenendaal, JianPRB}. The Hamiltonian includes the on-site Coulomb interaction between the 3d electrons and between the 3d electrons and the 2p core hole. The model parameters  are obtained within the Hartree-Fock limit and scaled down to 80\% to account for intra-atomic screening effects. The monopole parts were F$^0_{dd}$ = 6 eV and F$^0_{pd}$ = 7 eV. The spin-orbit coupling was included for the 3d and 2p electrons. The hybridization with the ligands was taken into account by including configurations up to a double ligand hole. The hybridization parameters used were  V = 2:25; -1:03 eV for the e$_g$ and t$_{2g}$ orbitals, respectively. The cubic crystal field of 10 Dq was set at 1.5 eV.


\textit{Results:} Figure 2(a) shows 2$\theta$ - $\omega$ scans around the (002) truncation rod for the 15 unit cell (uc) ENO films grown on different substrates. All samples show a broadened film peak (indicated by arrows), due to the reduced thickness of the films, and a sharp substrate peak which was used to align each data set. The film peaks for the highly compressive films (YAO and SLAO) show a noticeable shift from the bulk ENO lattice constant (represented by a dashed line) towards smaller 2$\theta$, while for LAO no such shift is observed. On the other hand, films grown under tensile strain (NGO and LGO) exhibit  a well resolved film peak with no measurable shift away from the bulk value. Figure 2(b) shows a reciprocal scan map (RSM) around the (-103) Bragg peak for a 35 uc ENO on NGO film (thicker films were necessary to resolve the peak using this conventional XRD). The weak film peak shares the same value for H (reciprocal lattice units) and a larger value for L, showing the film shares the in-plane lattice constant of the substrate.

After  the high structural  quality was established we turned our attention to  their transport properties shown in Fig 3(a). The data were recorded during both cooling and heating cycles from 380 K to 2 K; since no measurable hysteresis was found only the curves measured on warming are shown. As seen, for the tensile strain the resistivity follows the expected bulk-like insulating behavior below 380 K. This behavior, however, markedly changes after reversing the sign of strain. For the small compressive strain on  LAO, the resistivity at lower temperatures separates from bulk behavior at $\sim$ 250 K and begins increasing at a lesser rate. For the compressively strained film on SLAO, the sample shows unexpected metallic behavior at high temperatures with a MIT occurring at 335 K. And finally, for the largest  value of strain of -2.5 $\%$, the film on YAO turns metallic in the entire temperature range down to 2 K. To investigate the magnetic transition via the electrical transport, $\frac{dln\rho}{d(\frac{1}{T})}$ vs T is shown for all films besides ENO on YAO using a custom built liquid nitrogen cryostat to reduce measurement noise which is usually amplified by the derivative analysis, in order to mimick the analysis used by Zhou \textit{et al}\cite{Zhou}. Each film showed a characteristic kink indicative of an AFM transition around 200K. Fig. 3(c) shows the extracted T values for the kinks, denoted T*.

We performed measurements at  the Ni L$_{3,2}$ - edges using XAS to investigate the electronic structure of our films. As seen in Fig. 4a, all films show a strong white line at $\sim$ 855 eV  and $\sim$ 872 eV. Additionally, a shoulder around 853 eV  and 871 eV is apparent for all films, being much larger for the tensile case; this feature gradually decreases with increasing compressive strain. The size of the energy separation between the L$_3$ multiplet peak for each value of strain is plotted in Fig. 4(b) in the left axis. The splitting decreases from the tensile strained films, $\sim$ 1.8 eV, with increasing compressive strain to $\sim$ 1.2 eV for YAO. Fig. 4(b) also shows the calculated CT energy (right axis) which follows a very similar trend.


\textit{Discussion:} The large amount of epitaxial strain built into these materials is due to the extraordinary ability of the perovskite structural units to accommodate the strain through tilts/rotations and changes in lattice symmetry\cite{Jakprl}; it is these effects that ultimately lead to the observed modulation of the physical properties. Based on this, 2$\theta$ values, corresponding to the film peaks in Fig. 2(a), were used to calculate the out-of-plane lattice constants  yielding the following c-axis lattice constants: 3.86 \AA \emph{} (YAO), 3.84 \AA \emph{} (SLAO), 3.80 \AA \emph{} (NGO), and 3.81 \AA \emph{} (LGO), while for LAO strong overlap between the substrate peak and film peak due to the small strain value of -0.5\%, prevent a reliable c-axis lattice constant from being extracted. While for the samples with in-plane compressive strain the shift of the out-of-plane lattice constants from the bulk value of 3.80 \AA \emph{} is expected and consistent with tetragonal distortion of the unit cell, the samples under tensile strain show no significant shift (e.g. 0.26\% for ENO on LGO, much lower then the +2.5\% biaxial in-plane strain).

To ascertain whether or not this lack of c-axis lattice modulation was due to strain relaxation, unlikely in ultra-thin films, a RSM was taken to detect any deviation from epitaxial growth. The thicker sample (35 uc) was required in order to obtain a strong enough (-103) film  peak with the conventional source XRD. The H value of this film peak matches well with that of the substrate (demonstrated by the dotted line) confirming the film is fully coherent to the substrate (as was found for ENO on YAO\cite{growthpaper}), while the center L value of $\sim$ 3.053 r.l.u. gives an out-of-plane lattice constant of 3.793 \AA \emph{}, in excellent agreement with the rocking curve measurement (within 0.2 \%). This, along with the bulk-like c-axis lattice constant, implies the strain is compensated for by octahedral tilts and rotations, similar to that found for LaNiO$_3$ films under tensile strain\cite{Jakprl, May1}. Further work including  X-ray linear dichroism (XLD) measurements and density functional theory calculations need to be performed in order to confirm this. 

Tracking the evolution of the MIT with lattice modulation  revealed a very significant effect. As compressive strain is increased T$_{MIT}$ is gradually suppressed until entirely disappearing for the case of ENO on YAO. 
 In the case of intermediate compressive strain on SLAO, the  linear T-dependent metallic behavior is followed by a MIT shifted to 335 K, putting it very close to room temperature. 
 Unfortunately, the high temperature of the bulk MIT (480 K) prevents us from investigating the change in the T$_{MIT}$ for LAO, NGO, and LGO films. With the lack of hysteresis, characteristic of first order phase transitions, our results strongly imply that  epitaxial strain does not induce a first order transition in this material, as was seen in the bulk by application of  `chemical' pressure\cite{Frand}. Instead the results show that compressive strain acts to lower the transition temperatures akin to isotropic external pressure\cite{Lengsdorf, Zhou, Mazin}. The resistivity results are also strongly reminiscent of behavior of  ultra thin films of NdNiO$_3$, where it was proposed that a closing of the correlated gap is responsible for the quenching of the MIT by compressive strain\cite{JianNNO}.

The ultra-thin nature of the samples precludes direct investigation of  the sample by  way  of conventional magnetometry. In an attempt to locate the AFM transition temperature,  $\frac{dln\rho}{d(\frac{1}{T})}$ was extracted for all insulating samples; this analysis allowed Zhou \textit{et al} to  reveal a characteristic  spin ordering temperature in the bulk nickelates where spin ordering  appears as a kink in the T dependence\cite{Zhou}. As seen in Fig. 3(c), all the films exhibit a broadened kink around T$^{*}$. The error bars are meant to represent the approximate width of the kink, which is similar for all samples. The magnitude of  T$^{*}$ is approximately 12K lower for the tensile strained samples and  is similar to the change reported for external pressure\cite{Zhou}.
  For LAO, with very small compressive strain, the value is shifted upward to 203 K. For SLAO the value of T$^{*}$ is further shifted to 207 K, indicating that the Neel temperature for these films is shifted higher with higher value strain, analogous to the effect of external pressure reported by Zhou \textit{et al}\cite{Zhou}.  Resonant x-ray scattering measurements are needed in order to further confirm these changes in T$_N$ and investigate any possible changes in magnetic structure, which cannot be deduced  via transport.

Resonant soft X-ray  absorption has been extensively utilized in the study of ReNiO$_3$ perovskites\cite{MedardeXAS, JianNNO, Mizumaki}. The small thickness ($\sim$ 6 nm) of these films compares well with the probing depth ($\sim$ 12 nm) of TEY mode and allows us to explore the electronic structure of the entire sample. 
Figure 3 shows the measured absorption for the whole range of strain values. As seen, the strong white line at 855 eV and 872 eV corresponds to the L$_3$ (L$_2$) edge transition from the d$^7$: t$_{2g}^6$e$_g^1$ ground state to the \underline{c}t$_{2g}^6$e$_g^2$ (\underline{c} denotes a core hole) excited state. Another lower energy peak, $\sim$ 853 eV, which corresponds to the same electronic transition when strong electron localization is present, becoming prominent for the highly tensile strained samples. This systematic change in the multiplet / L$_3$ relative position and intensity holds valuable information about the hybridization of the d$^7$ and d$^8$\underline{L} states. The quantitative value of the observed splitting, which is simulated by tuning the charge transfer energy ($\Delta$) is plotted in Fig. 4(b)\cite{JianPRB}. In order to obtain the peak splitting the data was fitted with two Voight functions.  A direct inspection of the plot clearly shows that this splitting begins decreasing as the films are  compressed in the a-b plane, suggesting that a change in the degree of covalency between Ni and O is likely a cause of the observed transport properties. 
To confirm this, ab-initio cluster calculations were carried out using the charge transfer energy $\Delta$ as a a control parameter (see Fig. 4(b) (right side)). The results of the calculation directly  suggest that  changing $\Delta$ reproduces the observed splitting well, with a value as large as $\sim$ 2.5 eV for the tensile strained samples and being reduced down to the very  small value of $\sim$ 0.65 eV in the case of the all metallic film on YAO corresponding to an enhanced degree of covalency by almost four times (spectra can be seen in Ref. 52\cite{JianPRB}). 

The large reduction in $\Delta$ (approximately 1/4th of the saturated value for ENO on YAO) strongly implies that hybridization  is strongly increased between Ni-d and O-p orbitals.  In addition, the value of $\frac{dln\rho}{d(\frac{1}{T})}$ (Fig. 3(b)) at room temperature is approximately equal to the activation gap, showing the activation energy is steadily increased as compressive strain is reduced and nearly quenched for the tensile strained samples.
The reduction of $\Delta$ and the activation gap under compressive strain can be rationalized  in terms of changes in the Ni-O-Ni bond; as the lattice is compressed in plane an increase in the overlap of the O: p$_{x,y}$ and Ni: d$_{x^2-y^2}$ orbitals occurs. Furthermore, the increasing covalence is strongly resemblant of the effect of A-site cation exchange (Fig. 1)\cite{Medarde}. On the other hand, for the tensile strained samples,  the bond overlap would decrease, leading to a reduction in Ni-O hybridization, except for the fact that the splitting and transport appear to remain \textit{unchanged} with increasing tensile strain. This observation indicates that as the strain is changed from compressive to tensile the major d$^8$\underline{L} contribution  to  the ground state becomes effectively decoupled from the ionic d$^7$ state. It is interesting  to  note that these results are also compatible with the recent Density Functional Theory + Dynamical Mean Field Theory nickelate calculations, based on a site-selective Mott state, where the insulating gap is determined by the singlet formation energy between an O-p hole and Ni-d electron\cite{Park}; the theory also suggests the decoupling of the d$^7$ and the d$^8$\underline{L} states leading to an insulating ground state. Further corroborating our results, Wang \textit{et al} recently suggested that the insulating regime is largely controlled by the d-band occupancy, and not by the intra-electronic repulsion, which is strongly dependent on the charge transfer from oxygen ions\cite{Wang2}. 



To summarize, epitaxial ultra-thin films of ENO grown on a variety of substrates spanning both compressive and tensile strain were investigated with XRD, electric d.c. transport, resonant XAS, and first principle cluster calculations. The  sample's electronic properties were found to be highly tunable by strain, shifting the Mott transition tantalizingly close to room temperature and only affecting magnetic ordering transition T$_N$ to a small extent. The absence of  hysteresis  indicates that the transitions remains second-order, showing a key difference between epitaxial strain and A-site doping  implying that compressive strain effectively mimics external isotropic pressure. A combination  of XAS and ab-initio cluster calculations has determined that compressive strain enhances the covalence of the Ni-d and O-p orbitals, eventually leading to an entirely metallic ground state not  accessible in the bulk. These results showcase ENO's tunability, as the MIT can be tuned from the impractically high bulk value of 480 K to near room temperature or to being entirely quenched without the complication of chemical doping.

JC was supported by grants from  DOD-ARO (W911NF-11-1-0200). Work at the Advanced Photon Source is supported by the U.S. Department of Energy, Office of Science under grant No. DEAC02-06CH11357.

\clearpage

\begin{figure}[h]\vspace{-0pt}
\includegraphics[width=.8\textwidth]{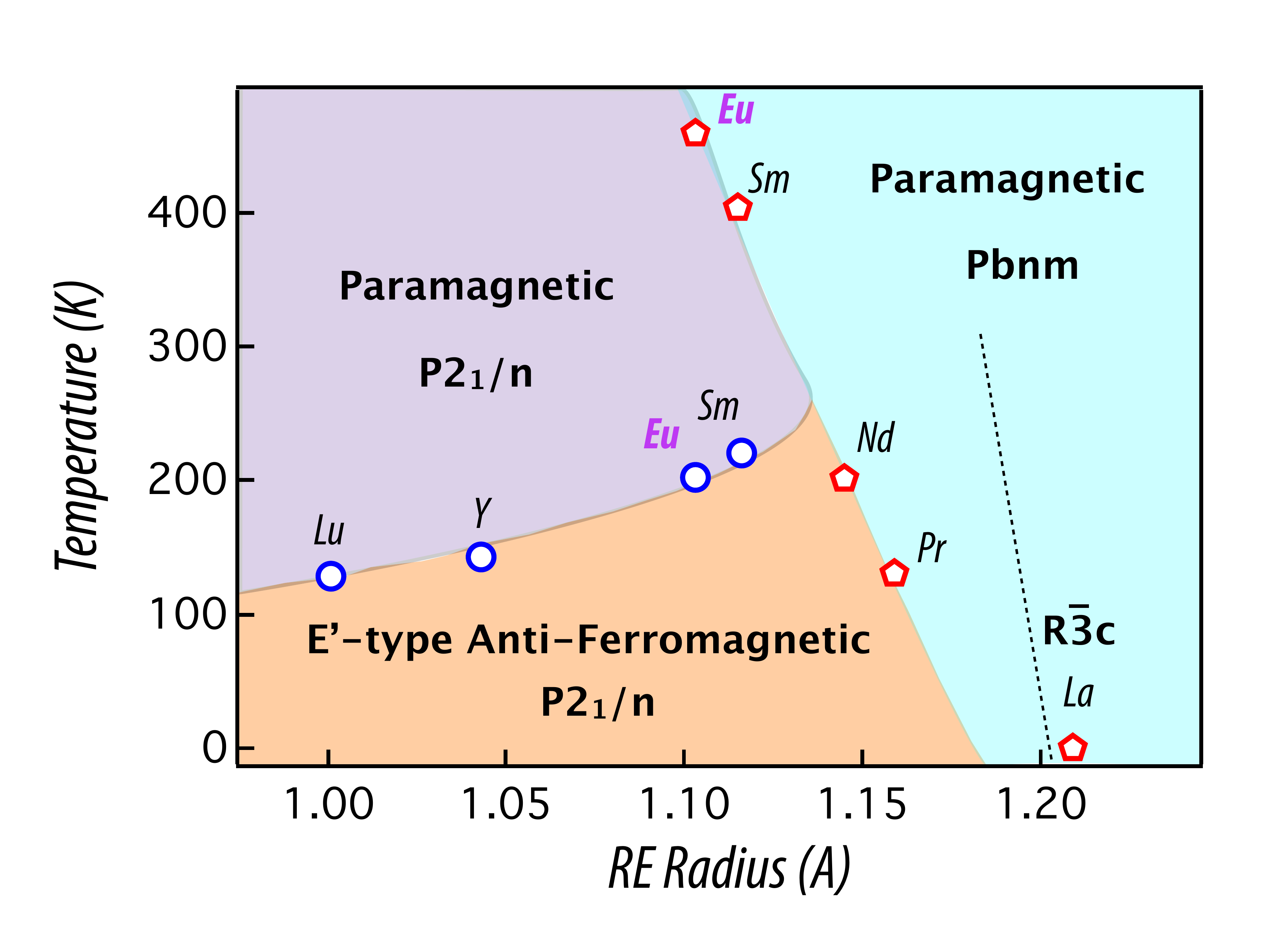}
\caption{\label{} (Color online) Partial phase diagram for the family of rare-earth nickelates of different A-sites (data  from Ref. 28\cite{Torrance}). The dotted line represents the change from orthorhombic (Pbnm) to rhomohedral (R$\overline{3}$C) symmetry.}
\end{figure}

\begin{figure}[h]\vspace{-0pt}
\includegraphics[width=.6\textwidth]{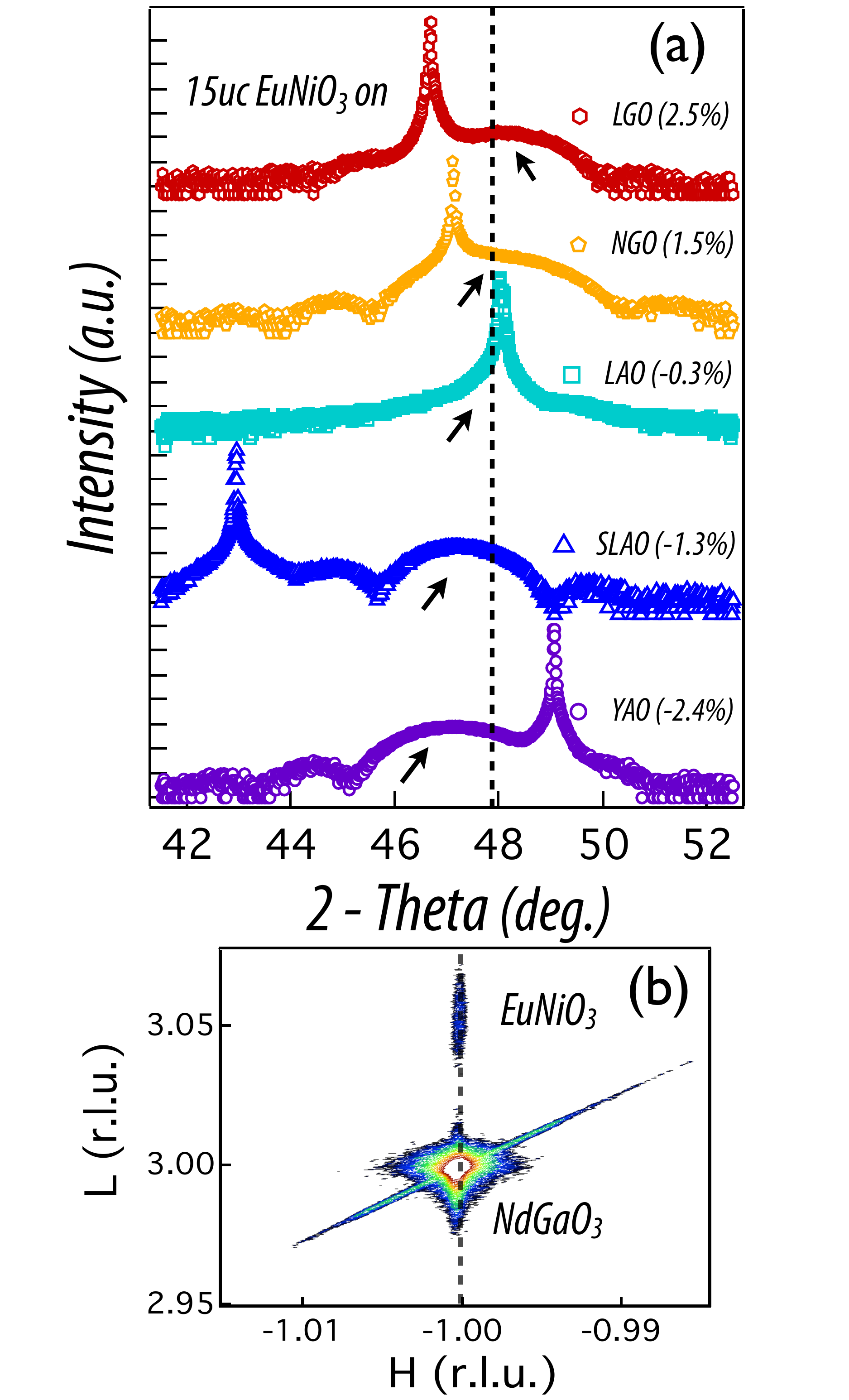}
\caption{\label{} (Color online) (a) XRD data for 15uc ENO samples on various substrates. The shifting from the bulk lattice value (indicated by the dashed line) is apparent for the compressively strained samples. The arrows indicate the film peaks. Note, for SLAO the (006) rod  is scanned due to the tetragonal structure. The data have been artificially shifted vertically to ease inspection. (b) RSM for a 35uc ENO film grown on NGO showing the film is coherently strained.}
\end{figure}

\begin{figure}[h]\vspace{-0pt}
\includegraphics[width=0.8\textwidth]{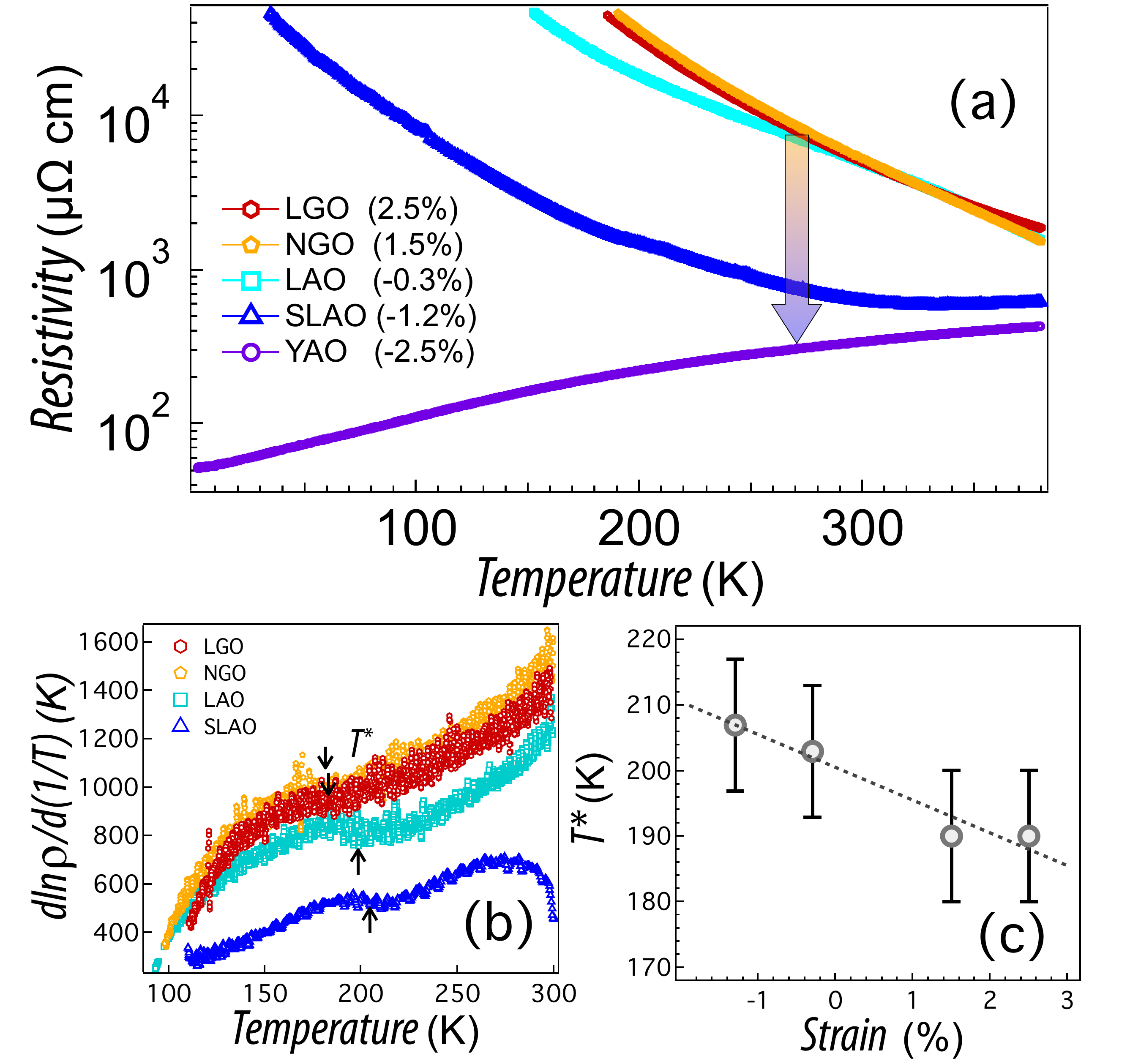}
\caption{\label{} (Color online) (a) Transport data for 15uc ENO samples on various substrates. The arrow indicates the direction of increasing compressive strain. (b) dln($\rho$)/d(1/T) data for the films. Arrows indicate the location of T*.  (c) T* for various strains.}
\end{figure}

\begin{figure}[h]\vspace{-0pt}
\includegraphics[width=0.8\textwidth]{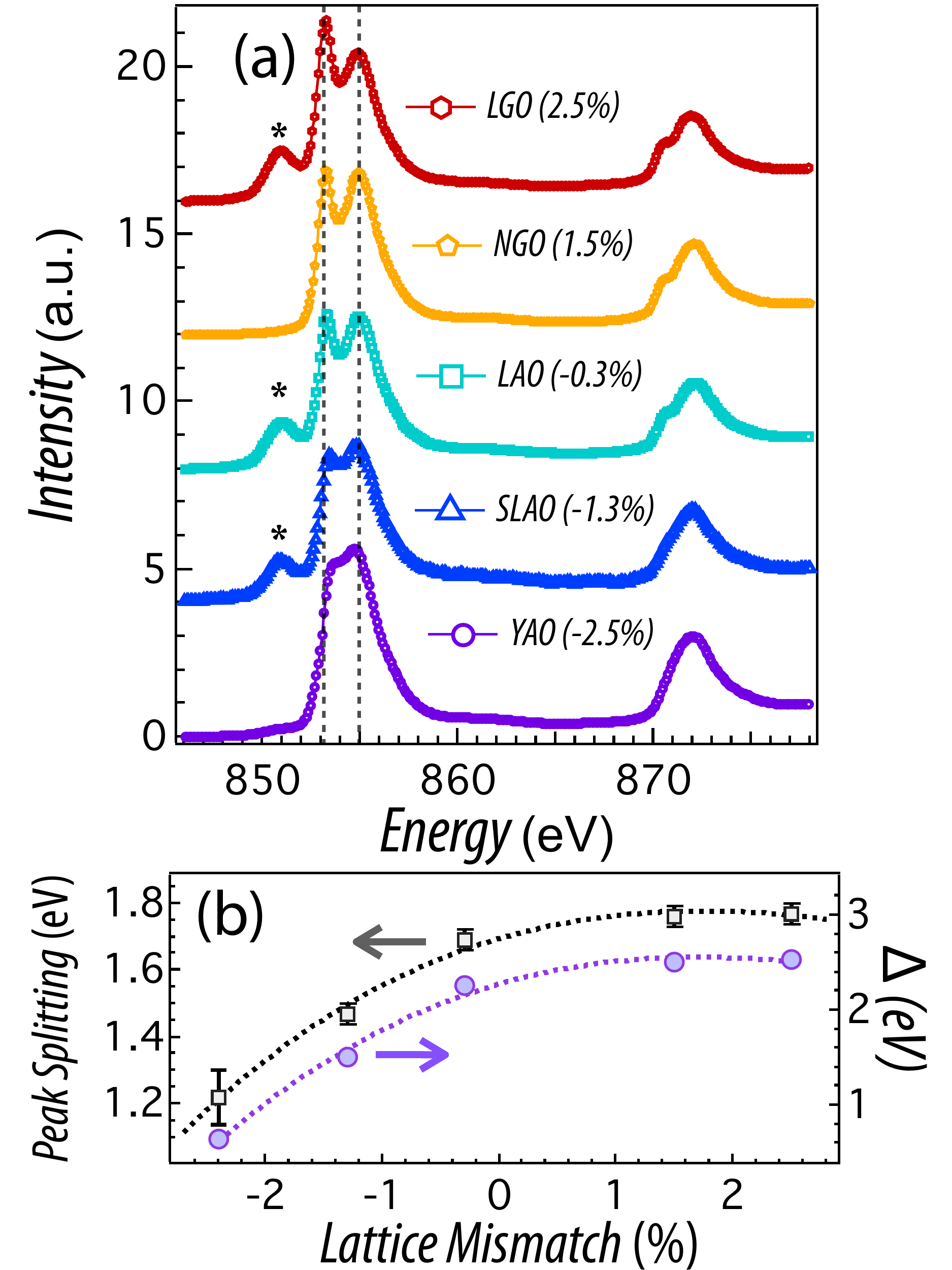}
\caption{\label{} (Color Online) (a) XAS data for 15uc ENO samples on various substrates showing the change in the multiplet peak with strain. The data have been artificially shifted vertically to ease inspection. (b) The experimentally obtained peak splitting along with the theoretically obtained corresponding CT energy. Note, a small peak, indicated by the asterisks, around $\sim$ 851 eV corresponding to the La M$_4$ edge appears for the substrates containing La.}
\end{figure}

\end{document}